\documentclass[prl,
                10pt,
                twocolumn,
                preprintnumbers,
                amsmath,
                amssymb,
                showpacs,
                superscriptaddress]{revtex4-1}

\usepackage[dvips]{graphicx}
\usepackage{dcolumn}  
\usepackage{bm}       
\usepackage{CJK}

\begin{document}

\begin{CJK*}{UTF8}{}

\title{Correlated dynamics of the motion of proton-hole wave-packets in a
       photoionized water cluster}

\author{Zheng Li \CJKfamily{gbsn}(李铮)}
\affiliation{%
Center for Free-Electron Laser Science, DESY,
            Notkestrasse 85, D-22607 Hamburg, Germany }

\author{Mohamed El-Amine Madjet}
\affiliation{%
Center for Free-Electron Laser Science, DESY,
            Notkestrasse 85, D-22607 Hamburg, Germany }

\author{Oriol Vendrell}
\email[]{oriol.vendrell@cfel.de}
\affiliation{%
Center for Free-Electron Laser Science, DESY,
            Notkestrasse 85, D-22607 Hamburg, Germany }

\author{Robin Santra}
\affiliation{%
Center for Free-Electron Laser Science, DESY,
            Notkestrasse 85, D-22607 Hamburg, Germany }
\affiliation{%
Department of Physics, University of Hamburg, D-20355 Hamburg, Germany}

\date{\today}






\begin{abstract}
    We explore the correlated dynamics of an electron-hole and a proton after ionization
    of a protonated water cluster by extreme ultra-violet (XUV) light.  An ultrafast
    decay mechanism is found in which the proton--hole dynamics after the ionization are
    driven by electrostatic repulsion and involve a strong coupling between the nuclear
    and electronic degrees of freedom. We describe the system by a quantum-dynamical
    approach and show that non-adiabatic effects are a key element of the mechanism by
    which electron and proton repel each other and become localized at opposite sides of
    the cluster. Based on the generality of the decay mechanism, similar effects may be
    expected for other ionized systems featuring hydrogen bonds.
\end{abstract}

\maketitle

\end{CJK*}

Correlated motions of electrons and protons in the so called proton coupled electron
transfer (PCET) reactions play a crucial role in charge transfer processes in
a range of biological, chemical and material science scenarios,
and have been the subject of intensive study over many
years~\cite{cukier_proton-coupled_1998,hammes-schiffer_theory_2010,weinberg_proton-coupled_2012}.
These reactions can be thermally activated processes occurring on relatively long time
scales, in which the reorganization of the molecular environment plays a decisive
role~\cite{cukier_proton-coupled_1998,hammes-schiffer_theory_2010}, as well as ultrafast
PCET processes occurring upon photoexcitation~\cite{li_ultrafast_2006}.
Some examples include the relaxation of electron vacancies in DNA and in crystals made of
DNA bases, which are known to proceed via correlated proton and electron transfer
reactions occurring in ps
time-scales~\cite{kumar_proton_2010,krivokapic_proton-coupled_2008}.
%
%
Proton transfer (PT) processes triggered by a sudden change in electronic structure are
also known to protect DNA base-pairs from photo-damage caused by UV-visible light
absorption and mediate the decay back to the ground electronic
state~\cite{sobolewski_tautomeric_2005}.
Non-adiabatic effects, in which the nuclear motion strongly couples to the evolution
of the electrons,
are ubiquitous in the kind of processes described above.

The response of matter to ionizing extreme ultraviolet (XUV) radiation in the range 20 to
100~eV has been the subject of much interest in the last years due to opportunities
offered by the generation of very short pulses in this energy range, opening the door to
time-resolved studies of the induced ultrafast dynamics in the fs time
regime~\cite{jiang_ultrafast_2010,popmintchev_attosecond_2010,rost_intense_2010,%
madjet_ultrafast_2011,farrell_strong_2011,kel11_043002,siu11_063412,zhou_probing_2012}.
The ionization from outer-valence shells can induce an ultrafast rearrangement of the
electronic cloud in what are referred to as hole migration
processes~\cite{kuleff_multielectron_2005,remacle_electronic_2006}.  The increased total
charge and the modified bonding structure of a sample may result in changes in its
nuclear configuration reaching possibly structures of strong non-adiabatic mixing between
electronic states.
Examples of ultrafast nuclear dynamics involving hydrogen atoms after extreme ultraviolet
(XUV) photoionization are known and have been measured by pump-probe techniques at the
Free Electron Laser at Hamburg (FLASH)~\cite{jiang_ultrafast_2010}.
The fragmentation of protonated water clusters H(H$_2$O)$_n^+$ after XUV irradiation and
removal of an outer-valence electron was recently measured at FLASH. The fragmentation
products and their kinetic energies were
recorded~\cite{lammich_fragmentation_2010,wolf_soft-x-ray_2010}.
Systems in which a proton, and thereby an extra positive charge, is supported by a number
of hydrogen-bonded molecules, are ubiquitous in soft condensed matter such as
water~\cite{marx_nature_1999} or in proton-exchange membranes such as those used in fuel
cell applications~\cite{kreuer_proton_1996}, where they mediate charge transport.
A time-resolved picture and most mechanistic details of the nuclear-electronic dynamics
after their ionization, are however not yet known.

Here, we investigate the decay and relaxation mechanism of the electron-hole generated by
absorption of XUV radiation in a prototypical water cluster, the so
called Zundel cation, H(H$_2$O)$_2^+$. We find an ultrafast coupled proton--hole
relaxation process operating after photoionization from the outer-valence shell.
The correlation function between the location of proton and hole shows that they respond
to each other within few fs and quickly separate irreversibly.
This is a consequence of the strong electrostatic repulsion between them, which
results in  potential
energy surfaces (PES) for the nuclear motion with large gradients, but this
is not the whole picture. As discussed below, strong non-adiabatic effects involving the
breakdown of the Born-Oppenheimer approximation play a crucial role in the relaxation
mechanism.
The fact that PES topologies are to a large extent related to the electrostatic repulsion
between the hole and the proton indicates a general class of vibrational-electronic decay
mechanisms operating in ionized hydrogen-bonded systems that should thereby be quite
insensitive to particular details of their molecular and electronic structure.

\begin{figure*}
   \includegraphics[width=\textwidth]{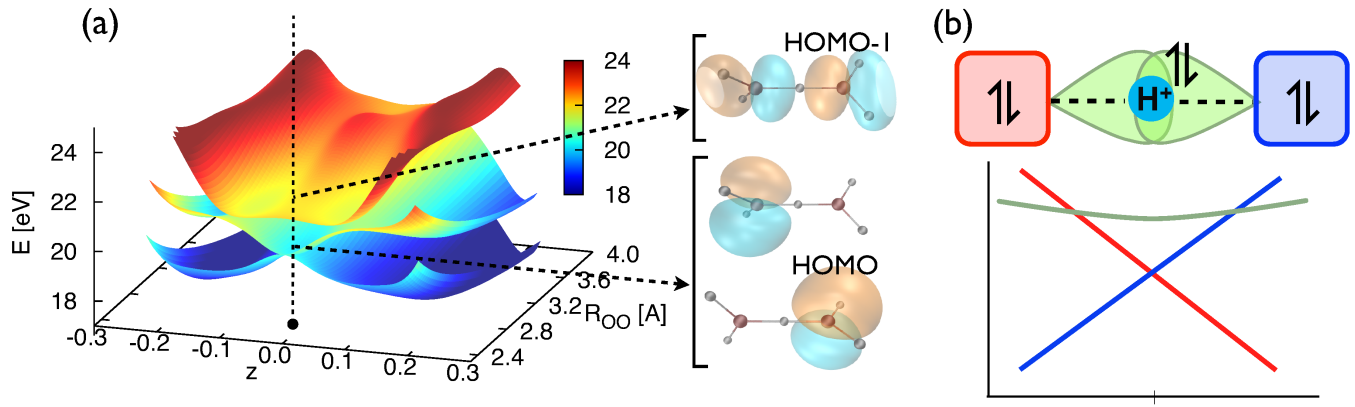}
     \caption{(color online) (a) Potential energy surfaces of the three lowest
         electronic states of H(H$_2$O)$_2^{++}$ for the $z$ (central proton position
         projected onto the oxygen-oxygen axis) and $R$ (oxygen-oxygen distance)
         coordinates and keeping the rest of vibrational modes frozen. The dot marks the
         Frank-Condon nuclear geometry.  The orbitals shown represent the highest
         occupied molecular orbitals at the Hartree-Fock mean field level, and they
         correspond to the electron-hole in the leading configuration for each of the
         three lowest energy electronic states of the dication in a multiconfigurational
         description. (b) Scheme of the hydrogen-bonded system indicating the location of
         the electron pairs that can be ionized and the corresponding potential energy
         surfaces as a function of the central proton position.  Ionization leading to
         the electronic states under consideration can take place from lone electron
         pairs on each monomer (in red and blue) or from the electrons involved in the
         hydrogen bonding (in green).
     } \label{fig1} \end{figure*}
\begin{figure}
   \includegraphics[width=8.6cm]{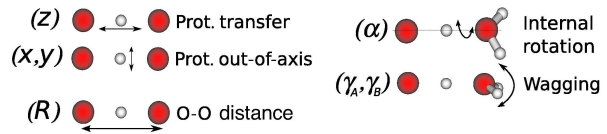}
     \caption{(color online). Schematic representation of the 7 internal nuclear
         coordinates used in the wavepacket propagations showing in each case the relevant
         atoms. The proton transfer coordinate used in this work was transformed as
         $z^{\prime} = z/(R-\delta)$~\cite{vendrell_full_2009}, where $\delta$ is a
     constant. Along this work appearances of $(z)$ refer to the $(z^{\prime})$ defined
     here.}
     \label{coords}
\end{figure}

The PES of the three lowest-energy states of the dication are shown in Fig.~\ref{fig1}a.
They were calculated at the complete active space self-consistent field level of
theory~\cite{olsen_determinant_1988} with the {\sc Molcas} suite of
programs~\cite{veryazov_2molcas_2004}.
The two nuclear coordinates of the 2D cut in Fig.~\ref{fig1}a correspond to the
oxygen-oxygen distance ($R$) and to the
projection of the central proton position onto the oxygen-oxygen axis ($z$),
whereas the rest of coordinates are held fixed.  A
schematic representation of the vibrational coordinates may be found in Fig.~\ref{coords}.
The point in coordinate space marked with a dot signals the position of the minimum
energy configuration in the ground electronic state of H(H$_2$O)$_2^+$. Therefore,
the nuclear wavepacket is centered at this position immediately after photoionization.
At such geometry of the cluster the two lowest electronic states of the dication are
energetically degenerate~\footnote{The two lowest electronic states of the dication
    belong to the $E$ irreducible representation within the D$_{2d}$ symmetry point
    group, which is adequate to classify the electronic states of the system at
Frank-Condon point}.  The orbitals shown in Fig.~\ref{fig1}a indicate the electron-hole
in a 1-particle picture. The electronic configurations with the missing electron in such
orbitals contribute to more than 90\% in the respective multiconfigurational electronic
wavefunctions, indicating that a 1-particle picture description of the states under
consideration is adequate in this case.
By inspecting them we realize that the two lowest-energy states can be described as
removing an electron from a lone electron pair in either water molecule.

The topology of the three low lying PES of the dication is illustrated schematically in
Fig.~\ref{fig1}b as a function of the central proton position.  The shape of the PES is
mostly the result of the electrostatic repulsion between the positive charge density
associated with the central proton and that of the electron-hole.  This is made clear by
noting that the gradient of the PES at the central point of conical intersection is
about $6.5$~eV$\cdot$\AA$^{-1}$.  This is equivalent to two point charges of equal sign
at a distance of $1.4$~\AA; the distance of the central proton to one of the
oxygen atoms is about $1.2$~{\AA} at the equilibrium geometry of the cluster.
When the central proton is displaced towards e.g. the water molecule on the left, the
electronic state with the hole on the water molecule on the right side stabilizes, while
the state with the hole on the left side destabilizes, resulting in the conical
intersection between the electronic states as a function of the proton position.
The third electronic state of the dication can be described by an electron-hole
delocalized mostly around the central proton. Therefore, the PES is quite flat along the
PT coordinate in the vicinity of the Franck-Condon point. When the proton is
displaced towards either water molecule, however, the electron-hole eventually becomes
more stable on a localized water site, and at that geometry the third dicationic state
crosses with the second one, resulting also in a seam of conical intersection.
%

The ultrafast dynamics of the protonated water dimer unfolding after outer-valence
ionization were computed quantum dynamically using the Heidelberg
implementation~\cite{meyer_mctdh_2007} of the multiconfiguration time-dependent Hartree
method~\cite{meyer_multi-configurational_1990,beck_multiconfiguration_2000}. Out
of the total $3N_a-6=15$ vibrational coordinates of the system, where $N_a$ equals the
number of atoms in the system, seven were kept active for the
dynamics calculations and the other eight were kept frozen at their ground state
equilibrium values. The active vibrational modes are shown in Fig.~\ref{coords} and
correspond to the three Cartesian coordinates
locating the central proton in the frame defined by the water molecules, the
oxygen-oxygen distance, the water molecule wagging modes, and the relative angle
between both water molecules around the axis connecting both oxygen atoms. The remaining
frozen coordinates correspond to the internal degrees of freedom of each monomer,
which are considered as rigid bodies, and the two rocking angles.
The separation in active and inactive modes is made based on the vibrational frequencies of
the involved motions. Higher frequency modes are kept frozen, since they are expected to
be less coupled to the geometric rearrangements after ionization. A detailed description
of the coordinates used and of the kinetic energy operator in this set of internal
coordinates may be found in Ref.~\cite{vendrell_full_2009}.

The Hamiltonian of the system in diabatic representation reads~\cite{worth_beyond_2004}
\begin{equation}
    \label{eq:diab}
    \hat{H} = \hat{T}_n \otimes \hat{\mathbf{1}}_d + \sum_{i,j} |i\rangle
    \hat{W}_{i,j}(\mathbf{Q}) \langle j|
\end{equation}
where $\hat{W}_{ij}$ operates on the space of the vibrational degrees of freedom,
$\hat{\mathbf{1}}_d=\sum_{i}|i\rangle\langle i|$ is the unit operator in the discrete
space of the diabatic electronic states, and $|i\rangle$ refers to the $i$th diabatic
electronic state. The diabatization procedure used is based on regularized diabatic
states~\cite{koppel_construction_2001} and details will be presented elsewhere.
After ionization, assuming a sudden electron ejection that leaves the system in the
$i$th electronic state, the corresponding wavepacket is
$\Psi_i(t=0)=\Phi_0(\mathbf{Q})\otimes|i\rangle$.  The nuclear part $\Phi_0(\mathbf{Q})$
corresponds to the ground vibrational state on the ground electronic state PES of
H(H$_2$O)$_2^+$.
Such wavepackets are propagated for every initial electronic state in coupled PES using
the Hamiltonian in Eq.~(\ref{eq:diab}). We assume the ionization cross-section to be
constant over the considered energy range for the initial electronic states. Therefore,
the observables calculated from each propagation are averaged with equal weight.

\begin{figure}[!tb]
\centering
   \includegraphics[width=8.6cm]{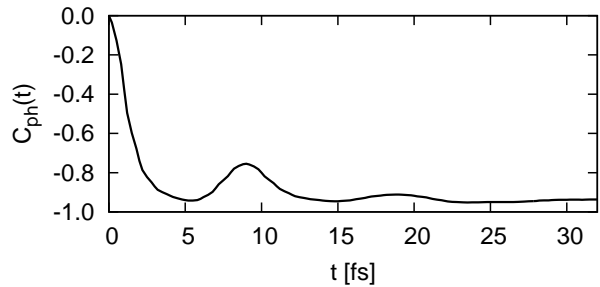}
     \caption{Correlation function between proton and electron-hole positions obtained
     from the propagated wavepacket.}
     \label{corr}
\end{figure}
The ultrafast correlated dynamics of hole and proton are made explicit by inspecting the
correlation function between their
positions, shown in Fig.~\ref{corr}. The correlation function is defined as
$C_{ph}=S_{ph}/\sqrt{S_{pp}S_{hh}}$, with
$S_{xy}=\langle\hat{x}\hat{y}\rangle-\langle\hat{x}\rangle\langle\hat{y}\rangle$.
$C_{ph}$ is bound to take values between $-1$, complete anti-correlation, and $+1$,
complete correlation.
For the excess proton position operator we simply take $\hat{z}$, the
projection of the central proton position onto the oxygen-oxygen axis of the cluster.  A
convenient definition of the hole position operator that can be readily applied to the
propagated wavepacket is
\begin{equation}
    \label{eq:corr}
   \hat{h}=\frac{1}{2}\hat{R}\otimes(|r\rangle\langle r| - |l\rangle\langle l|),
\end{equation}
where $|l\rangle$ and $|r\rangle$ refer to the two diabatic electronic states with the
hole localized on the left- and on the right-hand side water monomers, respectively,
and $\hat{R}$ is the position operator corresponding to the oxygen-oxygen distance.
According to this definition, the average location of the electron-hole corresponds to
the position of the left (right) oxygen atom when the system is in the electronic state
$|l\rangle$ ($|r\rangle$).
The correlation function in Fig.~\ref{corr} is obtained by summing over the different
initially populated electronic states.
At $t=0$ the proton and hole positions are still uncorrelated.  The degree of
anti-correlation increases very quickly as the proton is displaced towards one of the
water monomers and the hole moves in the opposite direction, reaching almost $-1$ in about
5~fs. A small recurrence in the correlation function is seen between 7 and 11~fs. It
originates from the bounce-back motion of the proton against the water monomers, which
have practically not yet started moving simply due to their higher inertia.
After 10 to 20~fs proton and hole have irreversibly localized at opposite sides of the
system, although the average oxygen-oxygen distance has only increased by 0.2~{\AA} and
larger recurrences could be expected. This irreversibility is a direct consequence of the
multidimensional nature of the triggered vibrational-electronic motion, which prevents
the wavepacket, just a few fs after photoexcitation, from returning to nuclear and
electronic configurations resembling the initially uncorrelated state.

\begin{figure*}
   \includegraphics[width=\textwidth]{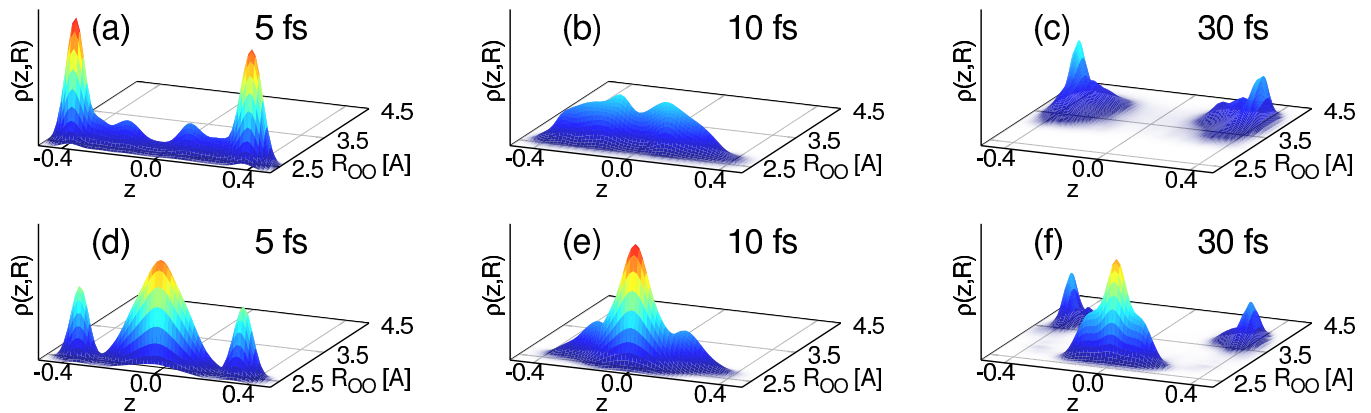}
     \caption{(color online). Reduced probability densities in the $z$ (central proton
         position projected onto the oxygen-oxygen axis) and $R$ (oxygen-oxygen distance)
         coordinates obtained from the propagated wavepackets after tracing over all other
         vibrational modes and electronic states. Snapshots (a), (b), and (c) correspond
     to propagations
  with the full
 Hamiltonian in Eq.~(\ref{eq:diab}).
 Snapshots (d), (e), and (f) correspond to propagation
 without non-adiabatic coupling in the Hamiltonian, Eq.~(\ref{eq:adiab}).}
     \label{dens}
\end{figure*}
The ultrafast nuclear dynamics and fragmentation after ionization are shown in
Figs.~\ref{dens}a, \ref{dens}b and \ref{dens}c, which
illustrate the dynamics of the cluster in the $z,R$ coordinates after integration over
all other nuclear coordinates and electronic states.
As the proton and hole quickly separate, the
central proton approaches the vicinity of an oxygen atom in less than 5~fs.  At about
10~fs there is a revival in which the proton bounces back to the center of the cluster
because, as already mentioned, the oxygen atoms did not yet have time to move.  After
15~fs the oxygen atoms have already separated by about 0.2~\AA, and the probability that
the proton is found at the central position between the water molecules has almost
vanished.  At this point, the correlated dynamics between the proton and the
electron-hole are over and they are found localized at opposite sites of the cluster,
leading to fragmentation of the system with 100\% probability.

In order to illustrate the dramatic role of the non-adiabatic coupling (NAC) between
nuclear motion and electron-hole dynamics, we briefly discuss the quantum dynamics of the
dication upon ionization in the adiabatic representation, but neglecting the NAC
altogether. Therefore the wavepackets evolve in uncoupled Born-Oppenheimer PES of the
ionized system and
the uncoupled Hamiltonian operator reads
\begin{equation}
    \label{eq:adiab}
    \hat{H} = \hat{T}_n \otimes \hat{\mathbf{1}}_a + \sum_{\alpha} |\alpha\rangle
    \hat{V}_{\alpha}(\mathbf{Q}) \langle \alpha|,
\end{equation}
where $\hat{\mathbf{1}}_a=\sum_\alpha|\alpha\rangle\langle\alpha|$ and
$|\alpha\rangle$ are now uncoupled adiabatic electronic states.
The reduced probability density $\rho(z,R)$ in which all other
nuclear coordinates and electronic states are integrated over is shown for this case
in Figures~\ref{dens}d, \ref{dens}e, and \ref{dens}f.
The part of the initial wavepacket on the ground electronic state PES of the dication,
i.e. initially below the lowest energy conical intersection, escapes from the
Franck-Condon region in about 5 to 10~fs, quickly leading to the dissociation channel
\mbox{H$_3$O$^+$ $+$ H$_2$O$^+$}. The reason is that in the lowest adiabatic electronic
state of the dication, the central proton and electron-hole are always at opposite sides
of the cluster (cf. Fig.~\ref{fig1}).
This is not the case when the nuclear dynamics start in the two lowest excited states of
the dication. Although the cluster is doubly charged, the missing coupling between nuclei
and electrons prevents it from fragmenting, the electron-hole and proton cannot rearrange
and the system stays bound. This is clearly seen by comparing Figs.~\ref{dens}a and
\ref{dens}d. It is worth mentioning at this point that the fragmentation experiments in
Ref.~\cite{lammich_fragmentation_2010} do not report unfragmented clusters after
ionization.

In conclusion, we have described an ultrafast decay and relaxation mechanism by which
an outer-valence hole generated by XUV light in a strongly hydrogen-bonded system
triggers proton--hole correlated dynamics unfolding in a time-scale of just a few fs.
The motion of the proton and the hole is anticorrelated and strongly coupled
and their
dynamics are driven by their electrostatic repulsion, which
results in PESs with large relative gradients at regions of intersection and strong
non-adiabatic effects. The rearrangement of the proton and hole
leads to electronic relaxation of the system, indicating that the dynamics
of the lightest nuclei is important when considering the motion of electron-holes
in molecular systems and clusters.
Artificially removing the non-adiabatic coupling in our quantum dynamical calculations
completely quenches the described mechanism because proton and hole cannot rearrange.
The electrostatic nature of the reported mechanism suggests that this class of decay
pathway may be a common feature of ionized hydrogen-bonded systems.

We thank the Helmholtz Association for financial support through
the Virtual Institute on ``Dynamic Pathways in Multidimensional Landscapes.''


\begin{thebibliography}{10}

\bibitem{cukier_proton-coupled_1998}
R.~I. Cukier and D.~G. Nocera,
\newblock Annu. Rev. Phys. Chem. {\bf 49}, 337 (1998).

\bibitem{hammes-schiffer_theory_2010}
S.~Hammes-Schiffer and A.~A. Stuchebrukhov,
\newblock Chem. Rev. {\bf 110}, 6939 (2010).

\bibitem{weinberg_proton-coupled_2012}
D.~R. Weinberg, C.~J. Gagliardi, J.~F. Hull, C.~F. Murphy, C.~A. Kent, B.~C.
  Westlake, A.~Paul, D.~H. Ess, D.~G. {McCafferty}, and T.~J. Meyer,
\newblock Chem. Rev. {\bf 112}, 4016 (2012).

\bibitem{li_ultrafast_2006}
B.~Li, J.~Zhao, K.~Onda, K.~D. Jordan, J.~Yang, and H.~Petek,
\newblock Science {\bf 311}, 1436 (2006).

\bibitem{kumar_proton_2010}
A.~Kumar and M.~D. Sevilla,
\newblock Chem. Rev. {\bf 110}, 7002 (2010).

\bibitem{krivokapic_proton-coupled_2008}
A.~Krivokapi{\'c}, J.~N. Herak, and E.~Sagstuen,
\newblock J. Phys. Chem. A {\bf 112}, 3597 (2008).

\bibitem{sobolewski_tautomeric_2005}
A.~L. Sobolewski, W.~Domcke, and C.~H{\"a}ttig,
\newblock Proc. Natl. Acad. Sci. {U.S.A.} {\bf 102}, 17903 (2005).

\bibitem{jiang_ultrafast_2010}
Y.~H. Jiang, A.~Rudenko, O.~Herrwerth, L.~Foucar, M.~Kurka, K.~U. K{\"u}hnel,
  M.~Lezius, M.~F. Kling, J.~van Tilborg, A.~Belkacem, K.~Ueda,
  S.~D{\"u}sterer, R.~Treusch, C.~D. Schr{\"o}ter, R.~Moshammer, and
  J.~Ullrich,
\newblock Phys. Rev. Lett. {\bf 105}, 263002 (2010).

\bibitem{popmintchev_attosecond_2010}
T.~Popmintchev, M.-C. Chen, P.~Arpin, M.~M. Murnane, and H.~C. Kapteyn,
\newblock Nat Photon {\bf 4}, 822 (2010).

\bibitem{rost_intense_2010}
H.~Chapman, J.~Ullrich, and J.~M. Rost,
\newblock J. Phys. B: At. Mol. Opt. Phys. {\bf 43}, 190201 (2010).

\bibitem{madjet_ultrafast_2011}
M.~E. Madjet, O.~Vendrell, and R.~Santra,
\newblock Phys. Rev. Lett. {\bf 107}, 263002 (2011).

\bibitem{farrell_strong_2011}
J.~P. Farrell, S.~Petretti, J.~Foerster, B.~K. {McFarland}, L.~S. Spector,
  Y.~V. Vanne, P.~Decleva, P.~H. Bucksbaum, A.~Saenz, and M.~Guehr,
\newblock Phys. Rev. Lett. {\bf 107} (2011).

\bibitem{kel11_043002}
F.~Kelkensberg, W.~Siu, J.~F. P{\'e}rez-Torres, F.~Morales, G.~Gademann,
  A.~Rouz{\'e}e, P.~Johnsson, M.~Lucchini, F.~Calegari, J.~L. {Sanz-Vicario},
  F.~Mart{\'\i}n, and M.~J.~J. Vrakking,
\newblock Phys. Rev. Lett. {\bf 107}, 043002 (2011).

\bibitem{siu11_063412}
W.~Siu, F.~Kelkensberg, G.~Gademann, A.~Rouzee, P.~Johnsson, D.~Dowek,
  M.~Lucchini, F.~Calegari, U.~De~Giovannini, A.~Rubio, R.~R. Lucchese,
  H.~Kono, F.~Lepine, and M.~J.~J. Vrakking,
\newblock Phys. Rev. A {\bf 84}, 063412 (2011).

\bibitem{zhou_probing_2012}
X.~Zhou, P.~Ranitovic, C.~W. Hogle, J.~H.~D. Eland, H.~C. Kapteyn, and M.~M.
  Murnane,
\newblock Nat. Phys. , 232 (2012).

\bibitem{kuleff_multielectron_2005}
A.~I. Kuleff, J.~Breidbach, and L.~S. Cederbaum,
\newblock J. Chem. Phys. {\bf 123}, 044111 (2005).

\bibitem{remacle_electronic_2006}
F.~Remacle and R.~D. Levine,
\newblock Proc. Nat. Acad. Sci. {\bf 103}, 6793 (2006).

\bibitem{lammich_fragmentation_2010}
L.~Lammich, C.~Domesle, B.~Jordon-Thaden, M.~F{\"o}rstel, T.~Arion, T.~Lischke,
  O.~Heber, S.~Klumpp, M.~Martins, N.~Guerassimova, R.~Treusch, J.~Ullrich,
  U.~Hergenhahn, H.~B. Pedersen, and A.~Wolf,
\newblock Phys. Rev. Lett. {\bf 105}, 253003 (2010).

\bibitem{wolf_soft-x-ray_2010}
A.~Wolf, H.~B. Pedersen, L.~Lammich, B.~Jordon-Thaden, S.~Altevogt, C.~Domesle,
  U.~Hergenhahn, M.~F{\"o}rstel, and O.~Heber,
\newblock J. Phys. B: At. Mol. Opt. Phys. {\bf 43}, 194007 (2010).

\bibitem{marx_nature_1999}
D.~Marx, M.~Tuckerman, J.~Hutter, and M.~Parrinello,
\newblock Nature {\bf 397}, 601 (1999).

\bibitem{kreuer_proton_1996}
K.-D. Kreuer,
\newblock Chem. Mater. {\bf 8}, 610 (1996).

\bibitem{olsen_determinant_1988}
J.~Olsen, B.~O. Roos, P.~J{\o}rgensen, and H.~J.~A. Jensen,
\newblock J. Chem. Phys. {\bf 89}, 2185 (1988).

\bibitem{veryazov_2molcas_2004}
V.~Veryazov, P.-O. Widmark, L.~Serrano-Andr{\'e}s, R.~Lindh, and B.~O. Roos,
\newblock Int. J. Quantum Chem. {\bf 100}, 626 (2004).

\bibitem{Note1}
The two lowest electronic states of the dication belong to the $E$ irreducible
  representation within the D$_{2d}$ symmetry point group, which is adequate to
  classify the electronic states of the system at Frank-Condon point.

\bibitem{vendrell_full_2009}
O.~Vendrell, M.~Brill, F.~Gatti, D.~Lauvergnat, and H.-D. Meyer,
\newblock J. Chem. Phys. {\bf 130}, 234305 (2009).

\bibitem{meyer_mctdh_2007}
H.-D. Meyer, G.~A. Worth, M.~H. Beck, A.~J{\"a}ckle, M.-C. Heitz, S.~Wefing,
  U.~Manthe, S.~Sukiasyan, A.~Raab, M.~Ehara, C.~Cattarius, F.~Gatti, F.~Otto,
  M.~Nest, A.~Markmann, M.~R. Brill, and O.~Vendrell,
\newblock {\em The {MCTDH} Package, Version 8.4, 2007.},
\newblock 2007,
\newblock http://mctdh.uni-hd.de.

\bibitem{meyer_multi-configurational_1990}
H.-D. Meyer, U.~Manthe, and L.~S. Cederbaum,
\newblock Chem. Phys. Lett. {\bf 165}, 73 (1990).

\bibitem{beck_multiconfiguration_2000}
M.~H. Beck, A.~J{\"a}ckle, G.~A. Worth, and H.-D. Meyer,
\newblock Phys. Rep. {\bf 324}, 1 (2000).

\bibitem{worth_beyond_2004}
G.~A. Worth and L.~S. Cederbaum,
\newblock Annu. Rev. Phys. Chem. {\bf 55}, 127 (2004).

\bibitem{koppel_construction_2001}
H.~K{\"o}ppel, J.~Gronki, and S.~Mahapatra,
\newblock J. Chem. Phys. {\bf 115}, 2377 (2001).

\end{thebibliography}

\end{document}